\documentstyle[12pt]{article}
\advance\textheight by \topskip
\begin{document}
\vspace{-0.2cm}
\begin{flushright}
SU-ITP-94-18\\
hep-th/9406093
\end{flushright}
\vspace{ 1 cm}

\begin{center}
{\large\bf DUAL WAVES}\\
\vskip 1.0 cm

{\bf Renata Kallosh} \footnote {  E-mail:
kallosh@physics.stanford.edu}
 \vskip 0.05cm
 Department of
Physics,
 Stanford University\\
Stanford   CA 94305, USA\\

\vskip 0.5cm
({\bf Contribution to Berezin Memorial volume})
\end{center}
\vskip 1 cm

\centerline{\bf ABSTRACT}
\begin{quotation}

We study the gravitational waves in the 10-dimensional target space of the
superstring theory. Some of these waves have unbroken supersymmetries. They
consist of Brinkmann metric and of a 2-form field. Sigma-model duality is
applied to such waves. The corresponding solutions we call dual partners of
gravitational waves, or dual waves. Some of these dual waves upon Kaluza-Klein
dimensional reduction to 4 dimensions become equivalent to the
conformo-stationary solutions of axion-dilaton gravity. Such solutions include
dilaton extreme black holes, axion-dilaton Israel-Werner-Perjes-type spacetimes
and extreme charged axion-dilaton Taub-Nut solutions. The unbroken
supersymmetry of the gravitational waves transfers to the unbroken
supersymmetry of axion-dilaton IWP solutions. More general supersymmetric
4-dimensional configurations derivable from 10-dimensional waves are described.

\end{quotation}

\newpage

\section{Introduction}

The fundamental reason for the existence of  any supersymmetric
non-renormalization theorem is related to the fact that  supersymmetric
theories  have a natural description in superspace, i.e. in a space with both
 bosonic
and fermionic coordinates. The famous rules of integration over anticommuting
variables were
discovered by Berezin. These rules form the basis for the proof of the generic
supersymmetric
non-renormalization theorems, which state that some semiclassical properties
of supersymmetric theories are not changed when
all quantum corrections
are taken into account.

In recent years an active field of research has been the search for
and also the study
of
non-perturbative solutions to the classical equations of motion of
superstring
effective
field theories and the corresponding sigma-models. Many bosonic
solutions of
such
theories have been discovered. Some special solutions turned out to
have a
highly
non-trivial property: although bosonic, they have some unbroken
supersymmetries.
This means that when embedded into the right supersymmetric theory
they admit
Killing spinors. Bosonic solutions with unbroken supersymmetries in
theories
of  quantum gravity are very special because they have some  kind of
supersymmetric non-renormalization property. Arguments that explain
these
properties were given for supersymmetric string solitons \cite{Da1}, \cite{CHS}
and
for extremal black holes \cite{US}.

A particularly interesting kind of metric is provided by the so
called {\it
pp-waves}\footnote{Here {\it pp-waves} stands for plane fronted waves
 with
parallel rays.
}.  Recently we have found the metrics in this class
which, together
with
appropriate dilaton, axion and gauge fields, provide solutions of the
lowest
order
superstring effective action and have unbroken supersymmetries \cite{BKO}. They
have been shown also to be free of   stringy
corrections.

There exists an extensive literature on gravitational waves.
We start by reviewing some results relevant for our purposes.
Pp-wave geometries are space-times admitting a covariantly
constant null vector field
\begin{equation}\label{null}
\nabla_{\mu} l_{\nu} = 0\ ,  \qquad l^{\nu}l_{\nu}= 0 \ .
\end{equation}
Spacetimes with this property were first discovered by Brinkmann  in
1923 \cite{Br1}.
In four dimensions the metrics of these spaces can be written in the
general
form
\begin{equation}\label{pl}
ds^2 = 2 du dv + K(u, \xi , \bar \xi ) du^2 -  d\xi d\bar \xi
\ ,
\end{equation}
where $u$ and $v$ are light-cone coordinates defined by
\begin{equation}
l_{\mu}=\partial_{\mu}u\ ,\hspace{3cm}l^{\mu}\partial_{\mu}v=1\ ,
\end{equation}
(thus the metric does not depend on $v$) and $\xi = x+iy$ and $ \bar
\xi =
x-iy$
are complex transverse coordinates. These metrics are classified and
described
in
detail in \cite{Kr1}. Different pp-wave spaces are characterized by
different
choices
of the function $K$ in eq. (\ref{pl}).  For example, when $K$ is
quadratic in
$\xi$ and $\bar{\xi}$,
\begin{equation}
 K(u,  \xi , \bar \xi ) = f(u) \xi^2 + \bar f(u) \bar
\xi^2 + g(u)
 \xi \bar  \xi \ .
\end{equation}
they are called {\it exact plane waves}.  Plane waves (\ref{pl}) with
$K$ of the form
\begin{equation}
K(u,  \xi , \bar \xi ) = \delta(u) f( \xi,  \bar  \xi) \ .
\end{equation}
are called {\it shock waves}.
A specific example of shock waves is given by  the  Aichelburg-Sexl
geometry
\cite{AS}
\begin{equation}
K(u,  \xi , \bar \xi ) = \delta(u) \ln \, ( \xi  \bar  \xi)\ .
\end{equation}
which describes the gravitational field of a point-like particle
boosted to the
speed of
light.

G\"{u}ven established in 1987 \cite{Gu1} that a solution to the
lowest order
superstring effective action, consisting of a
generalization of
the
four dimensional exact plane waves to $d=10$ with dilaton, axion and
Yang-Mills
fields, has  half of the $N=1, d=10$ supersymmetries unbroken and is
also a
solution of the equations of motion of the  superstring effective
action
including
all the $\alpha^\prime$-corrections. Investigations on the supersymmetry of
plane-fronted
waves
in general relativity were made even earlier. In \cite{Gi1} there is
a
reference to unpublished work of J.  Richer who found that pp-waves
are
supersymmetric. Very general classes of pp-waves were found by Tod to
be
supersymmetric in the context of $d=4, \,  N=2$ ungauged supergravity
in 1983
\cite{To1}. Different aspects of plane wave solutions in string
theory have
been
investigated in the  last few years \cite{Am1}, \cite{Ho1}.

The existence of a covariantly constant null vector field has
dramatic
consequences. For instance, for the class of d-dimensional pp-waves
with
metrics
of the form  \begin{equation}\label{d}
ds^2 = 2 du dv + K (u, x^i ) du^2 - dx^i d x^i \ ,
\end{equation}
where $i, j = 1,2, ..., d-2$, the Riemann curvature is \cite{Ho1}
\begin{equation}
R_{\mu\nu\rho\sigma} = - 2 l_{[\mu}( \partial_{\nu]}
\partial_{[\rho} K )
l_{\sigma]}\ .
\end{equation}
The curvature is orthogonal to $l_{\mu}$ in all its indices.
This fact is of crucial importance in establishing that all higher
order in
$\alpha^{\prime}$ terms in the equations of motion are zero due to
the
vanishing of all the
possible contractions of curvature tensors.
 G\"{u}ven  proved that the
corrections to the supersymmetry transformations  vanish.

In arbitrary dimension $d$ the most general metrics
admitting a covariantly constant null vector (\ref{null}),
were discovered by Brinkmann in 1925 \cite{Br2}:
\begin{equation}\label{BR}
ds^2 = 2 du dv + A_{\mu} (
u, x^i ) d x^{\mu} du - g_{ij}(
u, x^i )  dx^i d x^j \ ,  \quad l^\mu A_\mu =0\ ,
\end{equation}
where $\mu, \nu = 0,1, ...,  d-1$ and $i, j = 1,2, ..., d-2$.
Note that the
general
Brinkmann metric (\ref{BR}) in $d=10$ has 8 functions $A_i(u, x^i )$
and 28 functions $g_{ij}(u, x^i)$ more than the
metric (\ref{d}) investigated by G\"{u}ven,
where only the $uu$-component of the metric $A_u=\frac{1}{2} K$ was
present
and   was quadratic in $x^i$.

In \cite{BKO} we have found the 10-dimensional  supersymmetric
generalization of Brinkmann metrics.  We called this solution supersymmetric
string waves (SSW). The various properties of these solutions have been studied
recently \cite{BEK}, \cite{BKO2} \cite{KKOT}, \cite{BKO3}.
 In this paper we present a review of the recent results related to the dual
partners of supersymmetric string waves, which we call dual waves. In addition,
we describe shortly more general supersymmetric 4-dimensional
configurations, which
may be obtained from dual waves. Duality
transformations of string theory \cite{BUS}, which we are using here is
not a standard electro-magnetic duality which has been used in General
Relativity often. This is the reason why the gravitational waves are  usually
not related to to black-hole-type solutions.
 From the point of view of General Relativity those are
very different geometries. However the string theory brings in  a completely
different
concept of  ``equivalent''  background geometries. It was understood  some
time ago  \cite{Ho4} that the pp-waves are dual to the fundamental strings
\cite{Da1}.
The corresponding duality transformation, which is known as sigma-model
duality transformation  \cite{BUS},  has a very particular property: it changes
the value of the dilaton field $e^{2\phi}$ by the $g_{xx}$-component of the
metric, where $x$ is some direction on which all fields are independent. In our
recent paper \cite{BEK} we have established an analogous  dual relation between
 more general solutions of the effective equations of the critical ($d=10$)
superstring theories.

A remarkable property of a class of
SSW has been discovered recently \cite{BKO2}: their dual partners  are the
extreme 4-dimensional dilaton black holes \cite{G} upon Kaluza-Klein
dimensional reduction. And vice versa,
after $d=4$ extreme black holes are lifted up to $d=10$, the corresponding
configuration  became
a Brinkmann-type wave upon dual rotation.

In \cite{KKOT} we have found a general class of supersymmetric
solutions
of a 4-dimensional axion-dilaton gravity. Such
solutions include dilaton extreme black holes and axion-dilaton
Israel-Werner-Perjes-type spacetimes, which include also the
extreme charged axion-dilaton Taub-Nut solutions.
Here we will present the detailed derivation of these solutions from
SSW in $d=10$. This will require us to  to investigate a more general class of
SSW and to
establish their relation to a very general class of solutions of 4-dimensional
dilaton-axion gravity.

In this paper we will limit ourselves only  to
the zero
slope limit of the superstring effective action ($N=1, d=10$
supergravity). The coupling to
 Yang-Mills multiplet will enter via $\alpha^\prime$ corrections and  will
not be discussed here. It will be studied later along the lines  of
refs. \cite{BKO} and \cite{BEK}.

The paper is organized as follows. Sec.  2 we will describe supersymmetric
string waves and their dual partners.  We identify some of the dual waves
as the fundamental strings \cite{Da1} and indicate the relation
between some waves and 4-dimensional black holes.  Sec. 3 will describes some
details of
the supersymmetric Kaluza-Klein dimensional reduction of the bosonic part  of
the superstring effective action. Under such supersymmetric truncation the
unbroken supersymmetry
of 10-dimensional configuration is transferred automatically to unbroken
supersymmetry of the 4-dimensional configuration. In Sec. 4 we will perform
the supersymmetric Kaluza-Klein dimensional reduction of special
case of dual wave and we will compare it with 4-dimensional extreme black
holes.
In Sec. 5 we consider more general dual waves and
 compare it with 4-dimensional
stationary  solutions.
In Conclusion we will discuss this new method of generating
non-trivial
solutions  in gravitational theories and display new possibilities of deriving
4-dimensional supersymmetric solutions of $N=4$ supergravity interacting with
$N=4$ supersymmetric multiplets.

\section{Supersymmetric String Waves and the  Dual Waves }

We will use the sigma-model duality of the string theory and relate solutions
of 4-dimensional and 10-dimensional effective actions of string theory.
We will limit ourselves by keeping only one scalar field, the fundamental
dilaton. The pseudoscalar axion  field will appear in $d=4$ from the 3-form
field strength  $H$. The method which we develop here may give many other
interesting relations for the class of solutions which will include more
fields of the string theory.
Consider first the pp-waves \cite{Br1} in some dimension $d$.  The metric
is\footnote{Our notation are that of  \cite{BEK},
 \cite{BKO} and \cite{US}.}
\begin{equation}
ds^2 = 2dudv + 2A_u  du  du - \sum_{i=1}^{d-2} dx^i dx^i \ ,
\label{pp}\end{equation}
where the function $A_u$ depends only on transverse directions $x^i, \; i=1,
\dots , d-2$.
The equation that $A_u(x^i)$  has  to satisfy is:
\begin{equation}
\label{Lapl}
\triangle A_u = 0\  ,
\end{equation}
where the Laplacian is taken over the transverse directions only.
Sigma-model duality transformation \cite{BUS} defines the changes in the
metric, 2-form field $B_{\mu\nu}$ and in the dilaton field $e^{2\phi}$.
\begin{eqnarray}
 g_{xx} ' & =& 1/g_{xx}\ , \qquad  g_{x\alpha} ' =
B_{x\alpha}/
g_{xx}\ , \nonumber\\
  g_{\alpha\beta}' & =& g_{\alpha\beta} -
(g_{x\alpha}g_{x\beta} -
B_{x\alpha}B_{x\beta})/g_{xx}\ , \nonumber\\
  B_{x\alpha} ' & =& g_{x\alpha}/g_{xx}\ , \qquad
	  B_{\alpha\beta} ' = B_{\alpha\beta} +2
g_{x[\alpha}
B_{\beta]x}/g_{xx}\ , \nonumber\\
 \phi  '& =& \phi - {1\over 2} \log |g_{xx}| \ .
\label{bus}\end{eqnarray}
This transformation is defined for configurations with a non-null
Killing vector
in the $x$-direction.
The string theory considers such configurations as equivalent under
the condition
that the $x$-direction is compact. Leaving aside
this issue  we will start by presenting the upshot of  the dual relation
between
pp-waves (\ref{pp})
and fundamental strings \cite{Da1}.

Since we have chosen the metric of our pp-waves to be $u$-independent
and they are $v$-independent by the definition of  pp-waves, the metric
is also independent
 on $x= { 1\over \sqrt{2} }(u-v)$ and $t={ 1\over \sqrt{2} }(u+v) $. A
straightforward application of the sigma-model duality
transformations
given in (\ref{bus}) on pp-waves given in
eq.~(\ref{pp})
leads to the following new solution.
\begin{eqnarray}
\label{fs}
ds^2 &=&  2  e^{2 \phi}  dudv
 - \sum_{i=1}^{d-2} dx^idx^i\ ,\nonumber\\
B &=&  2 (1- e^{2 \phi})  du \wedge dv \ ,\\
e^{-2 \phi}  &=&1 -A_u \ .\nonumber
\end{eqnarray}
 Let the function $A_u$ be spherically symmetric and
depend only on $r^2 = \sum_{i=1}^{i=d-2} x^ix^i$. The choice
\begin{equation}
A_u = - { \mu \over r^{d-4}}
\end{equation}
 solves the harmonic equation (\ref{Lapl}) at  $r\neq 0$ if $\mu$ is a
constant.
With this choice of the function $A_u$ eq. (\ref{fs}) is the solution found
in \cite{Da1}  describing the field outside of a  fundamental string. For
$d=10$,  which is the critical dimension of the superstring theory, this
solutions looks as follows
\begin{eqnarray}
\label{fs2}
ds^2 &=&    (1 + { \mu \over r^{6}})^{-1}
  (dt ^2 - dx^2)
 -  \sum_{i=1}^{i=8} dx^idx^i\ ,\nonumber\\
\nonumber\\
B _{xt} &=&     { \mu \over r^{6} + \mu} \qquad
e^{-2 \phi}  =1 + { \mu \over r^{6}}
\ .
\end{eqnarray}
The dual partner of the fundamental string is the following pp-wave
\begin{equation}
ds^2 = 2 du dv  -   { 2 \mu \over r^{6}} du  du -  \sum_{i=1}^{i=8} dx^i dx^i \
,
\label{pp2}\end{equation}
and the 2-form field and the dilaton are absent (or could be equal to some
constants).
Note that in $d=10$ both the pp-waves as well as the fundamental strings have
unbroken supersymmetries \cite{Gu1}, \cite{Da1}.
With this reminder of the duality between the fundamental string and pp-wave
we
 may proceed to display the duality between supersymmetric string waves and the
``lifted'' extreme black holes.

Supersymmetric String Waves (SSW) \cite{BKO} are the  plane-wave-type
solutions of the
superstring effective action which have unbroken space-time
supersymmetries. They describe dilaton, axion and gauge fields in  a stringy
generalization of the  Brinkmann metric. Some conspiracy between the metric,
 the axion field and gauge fields is required.  We  will consider here only
the zero slope limit of the effective string action. This limit
corresponds to 10-dimensional $N=1$ supergravity. The Yang-Mills multiplet
will appear in the first order of $\alpha'$ string corrections.
The conspiracy between the metric
and the axion field is the characteristic property of the SSW and is necessary
to
provide unbroken supersymmetry even in the zero slope limit.

The SSW \cite{BKO} in $d=10$
 are given by the Brinkmann metric \cite{Br1}
and the following 2-form
\begin{eqnarray}
\label{SSW}
ds^2 &=& 2 d \tilde u d \tilde v + 2A_M d\tilde x^M d\tilde u -
 \sum_{i=1}^{i=8} d\tilde x^id \tilde x^i\ ,\nonumber \\
B &=& 2 A_M d\tilde x^M \wedge d\tilde u\ , \qquad
A_{v} =0 \ ,
\end{eqnarray}
where $i=1,\dots , 8, \; M = 0, 1, \dots , 8, 9$ and we are using the
 following notation for the
10-dimensional coordinates
$x^M = \{\tilde u, \tilde v, \tilde x^i\}$. We have put the tilde over
the 10-dimensional coordinates in this section, since we will have to compare
the original 10-dimensional
configuration with the 4-dimensional one, embedded into the 10-dimensional
space. A rather non-trivial identification of coordinates, describing these
solutions
will be required later.

Note that our SSW have a particular conspiracy between the metric
and the axion field
 \begin{equation}\label{choice}
g_{i \tilde u } = A_{i}= B_{i \tilde u }\ .
\end{equation}
  It is interesting that
 the solution in which the vector function in the metric
is related to the one in the axion was
mentioned by Tseytlin \cite{Ts1} as the most natural one from the point of
view of the sigma model
equations. In the work of Duval, Horv\'ath and Horv\'athy \cite{DHH} it was
found that such conspiracy between the metric
and the axion field leads to the absence of conformal anomaly at the
two-loop level.

The equations that $A_u(\tilde x^i)$ and $A_i(\tilde x^j)$ have to satisfy are:
\begin{equation}
\label{eq:Lapl}
\triangle A_u = 0\ , \hskip 1.5truecm \triangle\partial^{[i}A^{j]} =
0\ ,
\end{equation}
where the Laplacian is taken over the transverse directions only.
This solution has 8 functions $A_i$ more than that of pp-waves (\ref{pp}) where
only the function $A_u$ is non-vanishing.

A straightforward application of the sigma-model duality
transformations
given in (\ref{bus}) on the  SSW solution given in
eq.~(\ref{SSW})
leads to the following new supersymmetric solution of the
zero slope limit equations of motion:
\begin{eqnarray}
\label{dualwave}
ds^2 &=&  2e^{2\phi}\bigl \{ d\tilde ud\tilde v +  A_i d\tilde ud\tilde x^i
\bigr \} - \sum_{i=1}^{i=8} d\tilde x^id\tilde x^i\ ,\nonumber\\
B &=& -2 e^{2\phi} \bigl
\{ A_u d\tilde u \wedge d\tilde v +  A_id\tilde u \wedge d\tilde x^i \bigr \}\
,\\
e^{-2\phi}  &=& 1 - A_u \ ,\nonumber
\end{eqnarray}
where as before, the functions $A_{M} = \{A_u= A_u (\tilde x^j), A_v=0,
 A_i = A_i (\tilde x^j)\}$
satisfy equations (\ref{eq:Lapl}).
We call these solutions dual partners of the waves, or for simplicity, dual
waves.

We can make the following particular choice of the vector
function $A_M$. First of all  these functions will depend only on 3 of
the transverse coordinates, $\tilde x^1, \tilde x^2, \tilde x^3$, corresponding
to
our 3-dimensional
space. Secondly, we choose one of $A_i$ e. g. $A_4$ to be related to $A_u$.
\begin{equation}
A_u = - { \mu \over \rho}, \qquad  A_4 = \xi A_u \ , \qquad A_5 =
\dots = A_8=0 \ ,
\end{equation}
where $\rho^2 = \sum_{i=1}^{i=3} \tilde x^i  \tilde  x^i \equiv \vec x ^2$,
and $\mu$ is a constant.
We will specify the constant
$\xi$ later.
 Note that eqs. (\ref{eq:Lapl}) are solved outside $\rho=0$
\footnote{In order to solve the equations
(\ref{eq:Lapl}) everywhere,
it is understood that a source term at $\rho =0$, representing an unknown
object, perhaps a six-brane,
has to be added to these equations.  We hope that it can be  worked  out in an
analogy with the combined action for the macroscopic fundamental string,
where the source term comes from the sigma-model action, see eqs. (3,1) -
(3,3) in \cite{Da1}.}. We get
\begin{eqnarray}
ds^{2} & = &
2e^{2\phi}\{d\tilde{u}d\tilde{v}+\xi (1-  e^{-2\phi})d\tilde{x}^{4}d\tilde{u}\}
- \sum_{i=1}^{i=8} d\tilde{x}^{i}d\tilde{x}^{i}\, ,
\nonumber \\
B & = & -2e^{2\phi}\{(1-e^{-2\phi})\{d\tilde{u}\wedge
d\tilde{v}+ \xi d\tilde{u}\wedge d\tilde{x}^{4}\}\, ,
\nonumber \\
e^{-2\phi}  &=& 1 + { \mu \over \rho}
 \ .
\label{eq:new}\end{eqnarray}

 The solution given
in
(\ref{eq:new}) is different from  the solution of
\cite{Da1} corresponding to the field outside a fundamental string
and from our generalized FS solution, since it depends only on 3 coordinates
and not on 8. Also we take a particular case of our solution with relations
between $A_4$ and $A_u$.
We  perform the coordinate change
\begin{equation}
\hat{x}=\tilde{x}^{4} + \xi \tilde{u}\,  , \qquad
\hat  v = \tilde v  + \xi  \tilde x^4 \ .\end{equation}
We also shift $B$ on a constant value, since equations of motion depend on
$H=dB$ only.

 The dual wave solution (\ref{eq:new}) takes the form
\begin{eqnarray}
ds^{2} & = &
2e^{2\phi}d\tilde{u}d\hat {v}+\xi^{2}d\tilde{u}^{2}-d\hat{x}^{2}
-\sum_{i=1}^{i=3} d\tilde{x}^{i}d\tilde{x}^{i} -
\sum_{i=5}^{i=8} d\tilde{x}^{i}d\tilde{x}^{i}\,
, \nonumber \\
B & = & 2e^{2\phi}d\hat {v}\wedge d\tilde{u}\, ,
\nonumber \\
e^{-2\phi}  &=& 1 + { \mu \over \rho} \ .
\end{eqnarray}
When $\xi^2 = -1$ we have
\begin{eqnarray}
ds^{2} & = &
2e^{2\phi} d\hat{v} d\tilde{u} - d\tilde{u}^{2}-d\hat{x}^{2}
-\sum_{i=1}^{i=3} d\tilde{x}^{i}d\tilde{x}^{i} -
\sum_{i=5}^{i=8} d\tilde{x}^{i}d\tilde{x}^{i}\,
, \nonumber \\
B & = & 2e^{2\phi}d\hat{v}\wedge d\tilde{u}\, ,
\nonumber \\
e^{-2\phi}  &=& 1 + { \mu \over \rho} \ .
\label{partner}\end{eqnarray}
We can identify this particular dual partner of the SSW solution with
the uplifted dilaton black hole if we make the following identification
of coordinates
\begin{eqnarray}
t & = & \hat{v} =  \tilde v +\xi  \tilde x^4\, ,
\nonumber \\
x^{4} & = & \tilde{u}\, ,
\nonumber \\
x^{9} & = & \hat{x}= \tilde{x}^{4} + \xi \tilde{u}\, ,
\nonumber \\
x^{1,2,3,5,\dots,8} & =  & \tilde{x}^{1,2,3,5,\dots,8}\, .
\end{eqnarray}
Our dual wave becomes
\begin{eqnarray}
ds^{2} & = &
2e^{2\phi} dt d x^4 - \sum _4^9
d x^{i}d x^{i} -
d \vec x^2\,
, \nonumber \\
B & = & 2e^{2\phi}dt\wedge dx^4\, ,
\nonumber \\
e^{-2\phi}  &=& 1 + { \mu \over \rho} \ .
\label{10bh}\end{eqnarray}
This is an extreme electrically charged 4-dimensional black hole \cite{G},
which is
embedded into 10-dimensional
geometry in stringy frame, as we are going to explain in the next section.

\section{Supersymmetric dimensional reduction}

The embedding of the 4-dimensional bosonic solutions of the effective
superstring action into 10-dimensional geometry is not unique, in general
\footnote{We are grateful to E. Witten for  attracting our attention to
this problem.}. There are different ways to identify the vector field of the
charged black hole in 4 with the non-diagonal component of the metric
in the extra dimensions as well as with the 2-form field. Also the
identification of the 4-dimensional dilaton with the fundamental
10-dimensional dilaton and/or with some components of the metric in
the extra dimension is possible.

However the identification of  the 4-dimensional solution with the
10-dimensional one becomes  unique under the conditions
that the supersymmetric embedding for both solutions is identified.
Dimensional reduction of $N=1$ supergravity down to $d=4$ has
been studied by Chamseddine \cite{Cham} in canonical geometry. We
are working in stringy metric and also in slightly different notation.
In a subsequent publication we will present a detailed derivation of the
compactification of the bosonic part of the effective action of the
10-dimensional string theory
which is consistent with supersymmetry \cite{BKO2}.
Here we are interested in the relation between the  extreme dilaton
charged black
holes,
which have unbroken supersymmetry \cite{US} when imbedded into $d=4,
N=4$ supergravity\footnote{We do not know at present, whether the
embedding of these black holes into other theories, including the Abelian
part of Yang-Mills multiplet, will also correspond to some unbroken
supersymmetries.} and the corresponding 10-dimensional supersymmetric
configuration. The stationary IWP solutions of 4-dimensional
axion-dilaton gravity also happen to be supersymmetric when embedded into
$d=4, N=4$ supergravity. For for more general 4-dimensional supersymmetric
configurations, which have to be embedded into $d=4, N=4$ supergravity
interacting with $d=4, N=4$ supersymmetric matter multiplets, a more general
formulae will be required.

We start with the zero slope limit of the effective 10-dimensional
superstring action. The bosonic part of the action is
\begin{equation}
S=\frac{1}{2}\int d^{10}x
e^{-2 \phi}\sqrt{- g} \, [ -R+
4 (\partial\phi )^{2}-\frac{3}{4}H^{2}]\, ,
\label{eq:actionD1}
\end{equation}
where the 10-dimensional fields are the metric, the axion and the dilaton.

We want to make connection with the bosonic part of $N=4$, $d=4$ action.  In
this
particular case we are interested in compactifying $6$ space-like
coordinates. All fields are assumed to be independent of
six compactified dimensions. According to Chamseddine \cite{Cham}
dimensional
reduction of $N= 1, \, d=10$ supergravity to $d=4$
gives $N=4$ supergravity coupled to  6 matter multiplets. We are interested
here
only in dimensional reduction to $N=4$ supergravity without matter multiplets.

Let us first reduce from $d=10$ to $d=5$ by trivial dimensional reduction,
when we do not keep the non-diagonal
components of the metric and 2-form field.
We denote the
$10$-dimensional fields by un upper index ${}^{(10)}$ and the $5$-dimensional
fields by a hat.  The $10$-dimensional indices are capital letters
$M,N=0,\ldots,9$, the $5$-dimensional indices will carry a hat
$\hat{\mu},\hat{\nu}=0,\ldots,4$, and the compactified dimensions will
be denoted by capital $I$'s and $J$'s, $I,J=5,\ldots,9$.  We take the
$d=10$ fields to be related to the $d=5$ ones by
\begin{eqnarray}
g^{(10)}_{\hat{\mu}\hat{\nu}} & = & \hat{g}_{\hat{\mu}\hat{\nu}}\, ,
\nonumber \\
g^{(10)}_{I\hat{\nu}} & = & 0\, ,
\nonumber \\
g^{(10)}_{IJ} & = & \eta_{IJ}=-\delta_{IJ}\, ,
\nonumber \\
B^{(10)}_{\hat{\mu}\hat{\nu}} & = & \hat{B}_{\hat{\mu}\hat{\nu}}\, ,
\nonumber \\
B^{(10)}_{I\hat{\nu}} & = & 0\, ,
\nonumber \\
B^{(10)}_{IJ} & = & 0\, ,
\nonumber \\
\phi^{(10)} & = & \hat{\phi}\, .
\end{eqnarray}
We get
\begin{equation}
S=\frac{1}{2}\int d^{5}x e^{-2\hat{\phi}} \sqrt{-\hat{g}} [-\hat{R}+
4 (\partial\hat{\phi})^{2} -\frac{3}{4}\hat{H}^{2}]\, .
\end{equation}
As a second step we reduce from $d=5$ to $d=4$, keeping the non-diagonal
components of the metric and 2-form field. Since we are interested also in
supersymmetry, we will work with the 5-beins at this stage. The
4-dimensional indices
do not carry a hat.
We parametrize the
$5$-beins as follows
\begin{equation}
(\hat{e}_{\hat{\mu}}{}^{\hat{a}})=
\left(
\begin{array}{cc}
e_{\mu}{}^{a} &  A_{\mu} \\
0           &    1      \\
\end{array}
\right)
\, ,
\hspace{1cm}
(\hat{e}_{\hat{a}}{}^{\hat{\mu}})=
\left(
\begin{array}{cc}
e_{a}{}^{\mu} & -A_{a} \\
0           & 1   \\
\end{array}
\right)\, ,
\label{eq:basis}
\end{equation}
where  $A_{a}=e_{a}{}^{\mu}A_{\mu}$.
With this parametrization, the 5-dimensional fields decompose as follows
\begin{eqnarray}
\hat{g}_{44} & = & \hat{\eta}_{44}=-1 \, ,
\nonumber \\
\hat{g}_{4\mu} & = & - A_{\mu}\, ,
\nonumber \\
\hat{g}_{\mu\nu} & = & g_{\mu\nu}-A_{\mu} A_{\nu}\, ,
\nonumber \\
\hat{B}_{4\mu} & = & B_{\mu}\, ,
\nonumber \\
\hat{B}_{\mu\nu} & = & B_{\mu\nu}+A_{[\mu}B_{\nu]}\, ,
\nonumber \\
\hat{\phi} & = & \phi \ ,
\end{eqnarray}
where $\{ g_{\mu\nu},B_{\mu\nu},\phi,A_{\mu},B_{\mu} \}$ are the
$4$-dimensional fields.

The $4$-dimensional action for the $4$-dimensional fields becomes.
\begin{eqnarray}
S & = &
\frac{1}{2}\int d^{4}x e^{-2\phi}\sqrt{-g} [-R
+4(\partial\phi)^{2} -\frac{3}{4}H^{2}
\nonumber \\
&  &  +\frac{1}{4} F^{2}(A)
+\frac{1}{4} F^{2}(B)]\, ,
\end{eqnarray}
where
\begin{eqnarray}
F_{\mu\nu}(A) & = & 2\partial_{[\mu}A_{\nu]}\, ,
\nonumber \\
F_{\mu\nu}(B) & = & 2\partial_{[\mu}B_{\nu]}\, ,
\nonumber \\
H_{\mu\nu\rho} & = & \partial_{[\mu}B_{\nu\rho]}+
\frac{1}{2}\{A_{[\mu}F_{\nu\rho]}(B)+B_{[\mu}F_{\nu\rho]}(A)\}\, .
\end{eqnarray}
Now,  we study the dimensional reduction of gravitino. We are specifically
interested in  the supersymmetry transformation rule
of gravitino in $d=4$ supergravity without matter. This leads to
identification of
 the
matter vector fields $D_{\mu}$ and the supergravity vector fields
$V_{\mu}$.
\begin{eqnarray}
D_{\mu} & = & \frac{1}{2}(A_{\mu}-B_{\mu})\, ,
\nonumber \\
V_{\mu} & = & \frac{1}{2}(A_{\mu}+B_{\mu})\, ,
\label{vec} \end{eqnarray}
respectively.
 Now we want to truncate the theory
keeping only the supergravity vector field $V_{\mu}$. We have then
\begin{equation}
V_{\mu}=A_{\mu}=B_{\mu}\, ,
\hspace{1cm}
D_{\mu}=0\, .
\label{eq:truncation}
\end{equation}
The truncated action is\footnote{This action, which came from  the
10-dimensional
theory is slightly different from the
corresponding 4-dimensional action in our previous papers, e.g. in \cite{US},
due to
the difference in notation. The detailed explanation of this difference will be
given in \cite{BKO2}.}
\begin{equation}
S=\frac{1}{2}\int d^{4}x
e^{-2\phi}\sqrt{-g}[-R+4(\partial\phi)^{2}-\frac{3}{4}H^{2}
+\frac{1}{2}F^{2}(V)]\, ,
\label{eq:action4trunc}
\end{equation}
where
\begin{eqnarray}
F_{\mu\nu}(V) & = & 2\partial_{[\mu}V_{\nu]}\, ,
\nonumber \\
H_{\mu\nu\rho} & = & \partial_{[\mu}B_{\nu\rho]}+
V_{[\mu}F_{\nu\rho]}(V)\, .
\label{H}\end{eqnarray}

The embedding of the $4$-dimensional fields in this action in $d=10$ is
the following:
\begin{eqnarray}
g^{(10)}_{\mu\nu} & = & g_{\mu\nu}-V_{\mu}V_{\nu}\, ,
\nonumber \\
g^{(10)}_{4\nu} & = & - V_{\nu}\, ,
\nonumber \\
g^{(10)}_{44} & = & -1\, ,
\nonumber \\
g^{(10)}_{IJ} & = & \eta_{IJ}=-\delta_{IJ}\, ,
\nonumber \\
B^{(10)}_{\mu\nu} & = & B_{\mu\nu}\, ,
\nonumber \\
B^{(10)}_{4\nu} & = & V_{\nu}\, ,
\nonumber \\
\phi^{(10)} & = & \phi\, .
\label{eq:uplift}
\end{eqnarray}
This formulae can be used to uplift  any $U(1)$ $4$-dimensional field
configurations, including dilaton and axion,  to a $10$-dimensional field
configurations in a way consistent with supersymmetry.

The
conclusion of this supersymmetric dimensional reduction is the following.

i) The dilaton of the supersymmetric 4-dimensional extreme black holes is
identified as a fundamental
dilaton of string theory (and not one of the modulus fields).

ii)  Dimensional
reduction of $d=10$ supergravity to $d=4$
gives $N=4$ supergravity without 6 matter multiplets under condition that
$g_{4 \mu  }= - B_{4 \mu }$.  Therefore the
vector field of the 4-dimensional configuration is actually a non-diagonal
component of the metric in the extra
dimension as well as the 2-form field.
This works in our case since we have according to (\ref{10bh})
\begin{equation}
g_{4 t }^{(10)} = - B_{ 4 t }^{(10)} = -V_t = e^{2\phi}\ .
\end{equation}

iii) The supersymmetric truncation, presented above  has the following
important property by construction:  if the  4-dimensional configuration has
unbroken supersymmetries,
the uplifted one also has them and vice versa, if one starts with the
supersymmetric configuration in $d=10$ one ends up with the supersymmetric
configuration in $d=4$.
The reason for this is simple. The rules, given in eq. (\ref{eq:uplift}) have
been
derived
in such a way that the equations $\Psi = 0, \, \delta \Psi$=0  on all fermions
remain correct
and still have solutions with non-vanishing Killing spinors.

\section{Uplifting the Black Hole}
We will use the formulae from the section above to uplift the
dilaton black hole with one vector field.
The
electrically charged extreme 4d black hole is given by \cite{G} \footnote{There
is a
 difference of a $1/\sqrt{2}$ factor in the vector field with
respect to the one given in \cite{US}. }.
\begin{eqnarray}
ds^{2}_{str} & = & e^{4\phi}dt^{2}-d\vec{x}^{2}\, ,
\nonumber \\
V & = &  - e^{2\phi}dt\, ,
\nonumber \\
B & = & 0\, ,
\nonumber \\
e^{-2\phi} & = & 1+\frac{2M}{\rho}\, .
\end{eqnarray}
The uplifted configuration, according to eq. (\ref{eq:uplift}) is:
\begin{eqnarray}
ds^{2} & = & 2e^{2\phi}dtdx^{4}-d\vec{x}^{2}-(dx^{4})^{2}-
dx^{I}dx^{I}\, ,
\nonumber \\
B^{(10)} & \equiv & B^{(10)}_{MN}dx^{M}\wedge dx^{N}=- 2e^{2\phi}dx^4\wedge
dt \, ,
\nonumber \\
\phi^{(10)} & = & \phi\, .
\label{upl}\end{eqnarray}

Let us choose the parameter $\mu$ in the dual partner to the wave,
given in eq.  (\ref{partner}) equal to the double mass of the black hole.
\begin{equation}
\mu = 2 M\ .
\end{equation}
This makes the uplifted black hole (\ref{upl}) identical to the dual
partner to the wave,
given in eq.  (\ref{partner}).

For better understanding of black-hole-wave relation it is useful to do the
following.
By adding and subtracting from the metric the term
$e^{4\phi} dt^2$  we can rewrite  the dual wave in $d=10$, given
in eq. (\ref{10bh}) as follows.
\begin{eqnarray}
\label{eq:bh}
ds^2 &=&  e^{4\phi} d t ^2 -
 d \vec x^2 - (dx^4  -
 e^{2\phi} dt )^2
 -  dx^Idx^I
\ ,\nonumber\\
B &=&
  - 2e^{2\phi}dx^4\wedge
dt  \ ,\\
e^{-2\phi}  &=& 1 +  { 2M \over \rho}\ .\nonumber
\end{eqnarray}
 Now it is easy to recognize in the first 2 terms in the metric the
4-dimensional
metric and in the third term the non-diagonal component of the
10-dimensional metric  which together with the non-diagonal component of
the 2-form plays the role of the vector field in the 4-dimensional geometry.

 To show that our 10-dimensional solution is the embedding of the
extreme 4d black
hole we may also present the 10-dimensional metric
 in Kaluza-Klein  parametrization, where we have a 5d metric
$g_{\hat\mu \hat\nu} \times
x^I$-flat space
($\hat\mu = 0,1,2,3,4, I=5,\dots ,9$).
\begin{equation}
g_{\hat\mu \hat\nu} =
\left (\matrix{
0 & 0 & 0 & 0 & e^{2\phi} \cr
0 & -1 & 0 & 0 & 0\cr
0 & 0 & -1 & 0 & 0\cr
0 & 0 & 0 & -1 & 0\cr
e^{2\phi}  & 0 & 0 &  0 & -1\cr
}\right ) =
\left (\matrix{
g_{00} + A_0  A_0 g_{44} & 0 & 0 & 0 &  A_0 g_{44} \cr
0 & -1 & 0 & 0 & 0\cr
0 & 0 & -1 & 0 & 0\cr
0 & 0 & 0 & -1 & 0\cr
A_0 g_{44}   & 0 & 0 &  0 & g_{44}\cr
}\right )
\end{equation}
where $g_{00} = e^{4\phi} = (A_0 )^2, \;g_{11}=-1,
\;g_{22}=-1, \; g_{33}=-1, g_{44}=-1 $.
Thus the 10-dimensional non-diagonal component of the metric $g_{04}=
A_0 g_{44}$ together with the part of $B_{04}$ forms the 4-dimensional
vector field, as usual in Kaluza-Klein theory. The relation is, see eq.
(\ref{vec})
\begin{equation}
A_0 = - V_t  =  e^{2\phi} = {1\over 2 }
 ( g_{04}^{(10)} + B_{04}^{(10)}) \ .
\end{equation}
The left hand side of  this equation supplies the nice simple form of
the dual partner of the wave metric
\begin{equation}
ds^2 =  2 e^{2\phi} dt dx^4 -
d \vec x^2  - (dx^4)^2 - dx^Idx^I  ,
\end{equation}
the right-hand side shows that if this metric is rewritten in a more
complicated form by replacing the zero
in the upper right corner of the matrix by  $g_{00} + A_0
 A_0 g_{44} $,  we can recognize the $g_{00}$-piece of the
 4-dimensional black hole.

The case $\xi^2 = -1$ which gives  the 4-dimensional black hole in
Minkowski space with the signature  $(1,3)$
times the
compact 6-dimensional space with the signature $(0,6)$ corresponds to a
complex
10-dimensional wave in the space
with the signature  $(1,9)$.
\begin{eqnarray}
ds^2 &=& 2d\tilde u d \tilde v -  \; { 4M \over \rho}\;d\tilde
u( d\tilde u -i d\tilde x^4) - \sum_{i=1}^{i=8}
 d\tilde x^id\tilde x^i\ ,\nonumber \\
B &=& -i \;{ 4M \over \rho}\; d \tilde u \wedge d\tilde x^4\ .
\label{wave} \end{eqnarray}
By  performing a  rotation $i\tilde x^4 = \tilde \tau$  one can get
\begin{eqnarray}
ds^2 &=& 2d\tilde u d \tilde v  -  \; { 4M \over \rho }\;d\tilde u( d\tilde u
-d\tilde \tau)+ d\tilde \tau^2  -
\sum_{i=1}^{i=7} d\tilde x^i d \tilde x^i\ ,\nonumber \\
B &=&  - \;{ 4M \over \rho }\; d\tilde u \wedge d\tilde \tau\ \ .
\label{Brink}
\end{eqnarray}
This makes the wave real but with the signature of the space $(2,8)$.

Thus we may conclude that string theory considers as dual partners the
extreme 4d electrically
charged dilaton black hole  embedded into 10-dimensional geometry,
as given in eq. (\ref{eq:bh}) or (\ref{upl}),
and Brinkmann-type
10-dimensional  wave (\ref{wave}), (\ref{Brink}).

If we choose $\xi^2 = 1$ case we get the stringy equivalence between
Brinkmann-type 10-dimensional wave
\begin{eqnarray}
ds^2 &=& 2d\tilde ud \tilde v -  \; { 4M \over \rho}\;d\tilde
u( d\tilde u - d\tilde x^4) - \sum_{i=1}^{i=8}
 d\tilde x^id\tilde x^i\ ,\nonumber \\
B &=& -\;{ 4M \over \rho}\; d \tilde u \wedge d\tilde x^4\ ,
\label{SSW2} \end{eqnarray}
and lifted Euclidean 4-dimensional electrically charged dilaton
black hole with the signature
$(0,4)$ and
the 6-dimensional space has the signature  $(1,5)$,
\begin{eqnarray}
\label{eq:euclbh}
ds^2 &=& -e^{4\phi} d t ^2 -
d \vec x^2 + (dx^4 +
 e^{2\phi} dt)^2
- dx^Idx^I
\ ,\nonumber\\
B &=&
 - 2 e^{2\phi}dx^4 \wedge dt   \ ,\nonumber\\
e^{-2\phi}  &=& 1 +  { 2M \over \rho}\ .
\end{eqnarray}
With such choice of the signature the gravitational wave does not
have imaginary components. However,
the  fact that the metric as well as the 2-form field of the
gravitational wave in $d=10$ have an imaginary component  to be
dual to the lifted black hole in Minkowski space is strange. Note that this
is necessary only
if one insists that the $d=10$ space as well as the $d=4$ space are both
Minkowski spaces. One can avoid imaginary components by allowing
the changes in the signature of the space-time when performing duality
and dimensional reduction as explained above. Still this remains a puzzle.

 \section{From Waves to IWP and Vice Versa}

We consider again only  the  zero slop limit of supersymmetric string waves
(SSW)  \cite{BKO}
in $d=10$.
To generate a dual-wave, which upon dimensional reduction gives the
$4$-dimensional IWP solutions, we make the following choices for the
vector $A_M$ in the $10$-dimensional SSW (\ref{SSW}).  The components
$A_M$ are taken to depend only on three of the transverse coordinates,
$\tilde x^1, \tilde x^2, \tilde x^3$, which will ultimately correspond to our
3-dimensional space.  We choose one component, {\it e.g.}, $A_4$ to be
related to according to $A_4= \xi A_u$, where $\xi^2=\pm1$ depending on
the signature of spacetime.  Recall
that from (\ref{dualwave}) $A_u$ is
related to the dilaton by $A_u = 1-e^{-2\phi} $. For the remaining
components, we take only $A_1,A_2,A_3$ to be non-vanishing, and we
relabel these as $\omega_1,\omega_2,\omega_3$ for obvious reasons.  To
summarize, we then have
\begin{equation}
  A_4 = \xi A_u \ , \qquad A_1 \equiv \omega_1\ , \qquad A_2 \equiv
\omega_2\ , \qquad A_3 \equiv \omega_3\ ,
 \qquad A_5 = \dots = A_8=0.
\label{cond}\end{equation}
 We get  the following wave:
\begin{eqnarray}
ds^{2} & = &
2e^{2\phi}\{d\tilde{u}d\tilde{v}+\xi (1-  e^{-2\phi})d\tilde{x}^{4}d\tilde{u}
+ \omega_i d\tilde ud \tilde x^i \}
- \sum_{i=1}^{i=8} d\tilde{x}^{i}d\tilde{x}^{i}\, ,
\nonumber \\
B & = & -2e^{2\phi}\{(1-e^{-2\phi})\{d\tilde{u}\wedge
d\tilde{v}+ \xi d\tilde{u}\wedge d\tilde{x}^{4}  +
\omega_id\tilde u \wedge d \tilde x^i\}\, ,
\nonumber \\
e^{-2\phi}  &=& 1 + { \mu \over \rho}
 \ .
\label{new}\end{eqnarray}
The equations which the functions
$e^{-2\phi}$ and $\omega_i$ have to satisfy for the gravitational wave to
be a solution of equations of motion, following from the action
(\ref{eq:actionD1}),  are:
\begin{equation}
\triangle e^{-2\phi} = 0\ , \hskip 1.5truecm \triangle\partial^{[i}\omega^{j]}
=
0\ ,
\end{equation}
where again the Laplacian is taken over the transverse directions only.
These
equations
follow from eqs. (\ref{eq:Lapl}).

In what follows we repeat exactly the same steps which have allowed to show
that in absence of $\omega_i$ the dual partner of a wave is an extreme
black hole upon dimensional reduction.   We redefine $\tilde v$
 \begin{equation}
 \hat  v = \tilde v  + \xi  \tilde x^4 \ ,
\end{equation}
Solution takes the form
\begin{eqnarray}
\label{eq:new3}
ds^2 &=&  e^{2\phi} 2 d\tilde u (d \hat v +
\omega_i dx^i) - 2 \xi d\tilde u d\tilde x^4   - \sum_{i=1}^{i=8}(d\tilde
x^i)^2
\ ,\nonumber\\
B &=&
 2 (1-  e^{2\phi}) d\tilde u \wedge d \hat  v  -2 e^{2\phi}\omega_i
d\tilde u \wedge d \tilde x^i\ ,
\end{eqnarray}
The dilaton-axion field $\lambda$ satisfy the harmonic equation (\ref{harm}).
We may bring the metric to the form
\begin{equation}
ds^2 =  2 e^{2\phi} d\tilde u (d \hat v +
\omega_i d\tilde x^i) + \xi^2 d\tilde u^2   -
\sum_{i=1}^{i=3}d\tilde x^id \tilde x^i -
(d\hat x)^2 -
\sum_{i=5}^{i=8}d\tilde x^id \tilde x^i \ ,
\label{simple}\end{equation}
where \begin{equation}
\hat{x}=\tilde{x}^{4} + \xi \tilde{u}\ ,
\end{equation}
One can shift $B$ on a constant value, since equations of motion depend
on $H=dB$ only. We get
\begin{equation}
B =   2 e^{2\phi} (d \hat v + \omega_i d\tilde x^i) \wedge d \tilde u\ ,
\end{equation}

When $\xi^2 = -1$ we have
\begin{eqnarray}
\label{eq:bh1}
ds^2  &=&  2 e^{2\phi} d\tilde u (d \hat v +
\omega_i d\tilde x^i) - d\tilde u^2   -
\sum_{i=1}^{i=3}d\tilde x^id \tilde x^i -
(d\hat x)^2 -
\sum_{i=5}^{i=8}d\tilde x^id \tilde x^i
\ ,\nonumber\\
\nonumber\\
 B &=&   2 e^{2\phi} (d \hat v + \omega_i d\tilde x^i) \wedge d \tilde u
    \ ,\nonumber\\
\nonumber\\
 \triangle  \; e^{-2\phi}& =& 0\ ,  \qquad \triangle \;
\partial^{[i}\omega^{j]}=0 \ .
\end{eqnarray}

A particular identification of the
coordinates in the dual wave solution with those in the uplifted IWP
solution can be performed,
\begin{eqnarray}\label{dualwave2}
t & = & \hat{v} =  \tilde v +\xi  \tilde x^4\, ,
\nonumber \\
x^{4} & = & \tilde{u}\, ,
\nonumber \\
x^{9} & = & \hat{x}= \tilde{x}^{4} + \xi \tilde{u}\, ,
\nonumber \\
x^{1,2,3,5,\dots,8} & =  & \tilde{x}^{1,2,3,5,\dots,8}\, .
\end{eqnarray}
After all these steps our 10-dimensional dual wave becomes
\begin{eqnarray}
ds^{2} & = & 2e^{2\phi} d x^4 (dt   +\vec \omega \cdot d \vec  x) -
\sum_4^9 d x^{i}d x^{i} -d \vec x^2\, ,
\nonumber \\
B & = & - 2e^{2\phi} dx^4  \wedge  (dt +\vec \omega \cdot d \vec x) \, .
\end{eqnarray}

This allows to identify the  stationary supersymmetric axion-dilaton IWP
solution \cite{KKOT}, embedded into 10d geometry in stringy metric.
All arguments, given in  \cite{BKO2} about the uniqueness of the
embedding into higher dimension, once the
supersymmetry is identified, are valid in this case.

To recognize this as the lifted IWP solution, add and subtract from the
metric the term $e^{4\phi} (dt +\vec \omega \cdot d \vec x)^2$.  We can
then rewrite the dual wave metric (\ref{dualwave2}) as
\begin{equation}\label{eq:IWPwave}
ds^2 =  e^{4\phi} (dt +\vec \omega \cdot d \vec  x) ^2 -
 d \vec x^2 - \left(dx^4  -e^{2\phi} (dt + \vec \omega \cdot d
\vec x)\right)^2  -  \sum_5^9 d x^{i}d x^{i} \ .
\end{equation}
The first two terms now give the string metric for the $4$-dimensional
IWP solutions.  The non-diagonal components $g_{\mu 4}$, in the third
term, are interpreted as the $4$-dimensional gauge field components,
showing the Kaluza-Klein origin of the gauge field in this construction.
Note that the $4$-dimensional vector field components are also equal to
the off-diagonal components of the 2-form gauge field, giving the
overall identifications
\begin{equation}
g_{t 4 } = B_{t 4} =  e^{2\phi}= -V_t \ , \qquad g_{i 4} = B_{i 4}=
e^{2\phi} \omega_i= -V_i \ .
\end{equation}
The dilaton of the IWP solution, is identified with the fundamental
dilaton of string theory, rather than with one of the modulus fields.
The axion is identified with the $4$-dimensional part of the 3-form
field strength $H$ given in eq.  (\ref{H}).  Note that these
components of $H$ come totally from the second term in (\ref{H}),
since the 4-dimensional $B_{\mu\nu}$ vanishes.

To explain this relation better let us remind that the
stationary supersymmetric axion-dilaton IWP solution \cite{KKOT} is
\begin{eqnarray}
ds^{2}_{str} &=&e^{4\phi}(dt + \omega_i dx^i)^{2}- d\vec{x}^{2} \ ,
\nonumber\\
 \partial_{[i}\omega_{j]} &=& -{1\over 2} \epsilon_{ijk} \partial_k a,\qquad
A_{\mu}  = {1\over \sqrt 2} e^{2\phi}(1, \omega_i) ,  \nonumber\\
\triangle  \; e^{-2\phi}&=& 0\ , \qquad \triangle \; \partial^{[i}\omega^{j]}=0
\ .
\label{4d}\end{eqnarray}

The Kaluza-Klein parametrization of
the metric means the following relation between the higher-dimensional metric
$g_{MN}$ and the reduced one $g_{\mu\nu}$:
\begin{equation}
g_{MN} =
\left (\matrix{
g_{\mu\nu} + A_{\mu}{}^k g_{kl} A_{\nu}{}^l & A_{\mu}{}^k g_{ik} \cr
A_{\nu}{}^k g_{kj} & g_{ij}\cr }\right ) \ .
\end{equation}
 To show that our 10d
solution is the embedding of the axion-dilaton  4d IWP metric  we may present
the
10d metric
 in K.K. parametrization,
 where we have a 5d metric $g_{\hat\mu\hat \nu} \times
x^I $-flat space ($ \hat\mu, \, \hat \nu= 0,1,2,3,4)\quad I, \, J = 5, \dots ,
9$.

\begin{eqnarray}
g_{\hat\mu \hat \nu}  = \left (\matrix{
0 & 0 & 0 & 0 & e^{2\phi} \cr
0 & -1 & 0 & 0 & e^{2\phi}\omega_1\cr
0 & 0 & -1 & 0 & e^{2\phi}\omega_2\cr
0 & 0 & 0 & -1 & e^{2\phi}\omega_3\cr
e^{2\phi}  & e^{2\phi}\omega_1 & e^{2\phi}\omega_2 &
e^{2\phi}\omega_3 &
-1\cr }\right )=\qquad \qquad \qquad \qquad \nonumber\\
\nonumber\\
\nonumber\\
=\left (\matrix{
g_{00} + A_0{}^4  A_0{}^4 g_{44} & g_{10} +
A_1{}^4  A_0{}^4 g_{44}
& g_{20} + A_2{}^4  A_0{}^4 g_{44} & g_{01} +
 A_3{}^4  A_0{}^4
g_{44} &  A_0{}^4 g_{44} \cr g_{01} + A_0{}^4
A_1{}^4 g_{44} & -1 & 0 & 0
& A_1{}^4 g_{44}\cr g_{02} + A_0{}^4  A_2{}^4 g_{44} & 0 & -1 & 0 &
A_2{}^4 g_{44}\cr g_{03} + A_0{}^4  A_3{}^4 g_{44} & 0 & 0 & -
1 & A_3{}^4
g_{44}\cr A_0{}^4 g_{44}   & A_1{}^4 g_{44} & A_2{}^4 g_{44} &
 A_3{}^4 g_{44}
& g_{44}\cr }\right )
\label{matrix}\end{eqnarray}
where $g_{00} = e^{4\phi} = (A_0^4 )^2, \;g_{ii}=-1, g_{0i} =
e^{4\phi}
\omega_i$.  The 10-dimensional non-diagonal
components of the metric
$g_{04}=
 A_0^4 g_{44}, g_{0i} = A_{i}{}^4 g_{44}$ together with the part of $B_{04},
B_{0i}$ form
the 4d vector field, as usual in K.K. The relation is, see eq. (\ref{4d})
\begin{eqnarray}
A_0 = -V_t  =  e^{2\phi} = {1\over 2 } ( g_{04}
+ B_{04}) \ ,\nonumber\\
A_i =- V_i  =  e^{2\phi}\omega_i = {1\over 2 } ( g_{i4}
+ B_{i4}) \ .
\end{eqnarray}

The left hand side of   equation (\ref{matrix}) supplies the nice simple form
of
 the
dual partner of the wave metric
\begin{equation}
ds^2 =  2 e^{2\phi}  (d t + \omega_i dx^i)dx^4  - dx^idx^i- (dx^4)^2
 - dx^Idx^I\ .
\end{equation}
The right-hand side shows that if this metric is rewritten in a much more
complicated form by replacing many zero's
in the upper right corner of the matrix by  the terms like $g_{00} +
A_0^4  A_0^4 g_{44}, $ we can recognize the $g_{00}$-piece and other pieces of
the 4-dimensional axion-dilaton IWP metrics.

We may use
the 3-dimensional antisymmetric tensor $\epsilon_{ijk}$ to express
$\partial_{[i}\omega_{j]}$ through the axion field $a(x^i)$.
\begin{equation}
\partial_{[i}\omega_{j]} = -{1\over 2} \epsilon_{ijk} \partial_k a \ .
\end{equation}
By introducing the standard dilaton-axion complex field
$\lambda = a + i e^{-2\phi}$ we may rewrite equations (\ref{Lapl}) as follows:
\begin{equation}
\triangle \;\lambda = 0 \ .
\label{harm}\end{equation}

This accomplishes the derivation of the 4-dimensional IWP axion-dilaton
solutions
from the 10-dimensional gravitational waves.

\section{Conclusion}
Could we actually consider the dual relation between waves and
lifted black holes as something more than pure  algebraic curiosity?
 We believe that the answer to this question is ``yes''.  The dual relation
displayed above was established at the zero slope limit of the effective
action of the superstring theory. The issue of $\alpha' $-corrections in string
theory has
been studied extensively for the waves \cite{BKO}, \cite {BEK}. The pp-waves
have the best known
properties of absence of such quantum corrections \cite {Gu1}. The SSW  are
known to
have special property of the absence of $\alpha' $-corrections under the
condition that the non-abelian Yang-Mills fields  is added to the configuration
which at the zero slope limit $\alpha' =0$ consists only of the metric and
2-form \cite{BKO}.

It was explained in \cite{BEK} that the importance of
sigma-model duality  between supersymmetric configurations is in the
fact that the structure of
$\alpha'$ corrections is under control for the
dual solution if it was under control for the original solution.
In this way we have found  that the nice properties of the pp-waves \cite{Gu1}
are
carried over to the fundamental string solutions \footnote{ We have found
in    \cite{BEK}  that
under natural assumptions the fundamental strings are exact solutions of string
theory due to unbroken supersymmetry and dual relations to gravitational waves.
The exactness of fundamental string solutions was confirmed recently
by Horowitz and Tseytlin \cite{Ho5} by the analysis of the $\beta$-functions
and specific renormalization scheme in conformal field theory.}.
The present investigation
shows that
the electrically charged extreme black hole embedded into ten-dimensional
geometry may require to be supplemented by some non-abelian Yang-Mills
configuration, to avoid the possible $\alpha' $-corrections. In this respect
we would like to stress that the study of the properties of quantum corrections
established via duality may become a powerful mechanism of the
 investigation of  quantum theory despite the strange imaginary factors
in the waves, which are dual partners of the uplifted black holes.

At the very minimal level one can consider the method developed above,
which consists of stringy duality combined with Kaluza-Klein dimensional
reduction, as the
solution generating method. This method has the advantage of generating new
supersymmetric solutions from the original ones. If  we would not know that
 extreme
4-dimensional black holes are supersymmetric, we would discover this
via the supersymmetric properties of 10-dimensional waves.
If we would not found the IWP axion-dilaton spaces  in the 4-dimensional world,
we could have found them from SSW via dual rotation and dimensional reduction.
Simultaneously we would establish the existence of unbroken supersymmetry of
IWP geometries.

The SSW solutions found in \cite{BKO} are not the most general supersymmetric
wave solutions, more general ones may exist. However, if we look only on those
SSW solutions, about which we know that they are supersymmetric, could we still
use them to produce more general 4-dimensional configurations with unbroken
supersymmetries? The answer to this question is positive. Indeed, we may relax
few conditions, used in eq. (\ref{cond}): we may consider $A_4$ unrelated to
$A_u$ and choose $A_5, \dots , A_9$ all non-vanishing. The net result of this
is
the following. The fields which become part of the six vector matter
supermultiplet upon dimensional reduction of dual waves are non-vanishing
for new solutions. Indeed, by choosing $A_4$ to be proportional to $A_u$
we have achieved that the metric in $x^4$ direction
is flat. Otherwise, $g_{44}$ would not be equal to a constant. The choice
$A_5 \neq 0$ will make the
6-dimensional space non-flat,  $ g_{45}=- B_{45} = e^{2\phi} A_5 $, etc. The
combination  of vector  fields
 which enter the matter multiplet will be non-vanishing for these solutions,
since the
definition of vector fields will include above mentioned $g_{4I}$-component of
the metric and $B_{4I}$-component of the  2-form.
The corresponding
4-dimensional theory upon dimensional reduction will contain additional vectors
and scalars as well as pseudoscalars.  By
construction, we will have a bosonic action and configurations which solve
its equation of motion. These configurations will have half of unbroken
supersymmetries when embedded into $N=4$ supergravity  with six
abelian vector multiplets. We can also  take into account the Yang-Mills
part of the 10-dimensional gravitational wave solution SSW. This will lead to
the
 4-dimensional  solutions of equations of motion which have unbroken
supersymmetry when embedded into $N=4$ supergravity  with six
abelian vector supermultiplets  and non-abelian Yang-Mills
supermultiplet. The action and the corresponding
solutions will be presented in an explicit form in the future publication.

I am very grateful to my collaborators E. Bergshoeff, D. Kastor and  T. Ort\'\i
n
who made it possible to establish some deep properties of the gravitational
waves
in string theory and their  relation to the most interesting  4-dimensional
 geometries, including extreme black holes.

This work  was  supported by  NSF grant PHY-8612280.

\end{document}